# Post-earthquake modelling of transportation networks using an agent-based model

Kairui Feng[1], Quanwang Li[1] and Bruce R. Ellingwood[2]
[1]Department of Civil Engineering, Tsinghua University, Beijing, China
[2]TNList, Tsinghua University, Beijing China; Department of Civil and Environmental Engineering, Colorado State University, Ft. Collins, CO, USA

**Abstract**

Surface transportation systems are an essential part of urban transportation infrastructure and are susceptible to damage from earthquakes. This damage, along with the lack of prior warning of earthquake events, may lead to severe and unexpected disruption of normal traffic patterns, which may seriously impair post-disaster response. Accordingly, it is important to understand and model the performance of urban transportation systems immediately following an earthquake, to evaluate its capability to support emergency response, e.g., the movement of firefighters, search and rescue teams and medical personnel, and the transportation of injured people to emergency treatment facilities. For this purpose, a scenario-based methodology is developed to model the performance of a transportation network immediately following an earthquake using an agent-based model. This methodology accounts for the abrupt changes in destination, irrational behavior of drivers, unavailability of traffic information and the impairment to traffic capacity due to bridge damage and building debris. An illustration using the road network of Tangshan City, China shows that the method can capture the traffic flow characteristics immediately after an earthquake and can determine the capability of the transportation network to transfer injured people to hospitals considering the above factors. Thus, it can provide rational support for evaluating the performance of the surface transportation system under immediate post-disaster emergency conditions.

# 1 Introduction

Transportation systems are a major part of public civil infrastructure, and are essential for the normal functioning of society. Under emergency conditions, e.g., immediately following an earthquake, they are also needed for the movement of firefighters, search and rescue personnel and medical teams, and for transferring injured people to emergency treatment or healthcare facilities. As a result of increased traffic demand and impaired traffic capacity following the earthquake, traffic flows may be severely disrupted or become significantly heavier in some links of the transportation network (Shen et al., 2008). It is essential that transportation systems remain operational following a disaster so that the movement of emergency response vehicles is not impeded by severe congestion and excessive delay.

Many studies have considered earthquake-resistant design and retrofit of transportation infrastructure (Priestley et al., 1996; Winget et al., 2005). However, the disruptive effects of earthquakes depend not only on the extent of seismic damage to transportation infrastructure but also on the functional impairment of the transportation network as a result of that damage. Thus, it is important that the transportation system remains operational or its functions recover quickly in the immediate aftermath of the earthquake. Traffic flow may be disrupted significantly by a severe earthquake and become heavier because many travelers may change their routes and additional travel demands are placed by the movement of emergency response teams. For example, the $M_w$ 6.2 Christchurch, New Zealand earthquake of 2011, causing 185 fatalities and approximately 8,600 injuries, led to high traffic congestion throughout the Christchurch transportation network (Jacques, et al., 2014). Therefore, conventional traffic assignment methods are unsuitable for post-earthquake situations because they estimate the traffic patterns by assuming that the origin-destination (OD) demand is uniformly distributed over time. In the immediate aftermath of a severe earthquake, a dynamic traffic assignment (DTA) is necessary to predict the traffic patterns and formation of congestion in the transportation network immediately after its perturbation.

Considerable research has been dedicated to the development of the DTA method (Merchant and Nemhauser, 1978; Szeto et al., 2011). Chang et al. (2012) compared the user equilibrium (UE) model with the DTA model in post-earthquake modelling of traffic flow, and found that the DTA model captures the formation of congestion more accurately, while the UE model may generate unreasonable results, e.g., the total system travel time is underestimated by approximately 10 times. Miller and Baker (2016) employed an activity-based model to obtain the ODs and traffic mode assignment in the post-earthquake traffic modelling. However, it is assumed in the DTA that the drivers come to an equilibrium spontaneously and immediately under any given transportation network. This assumption may be suitable for modeling the weekly or monthly variation of traffic patterns (Tanimoto, 2015), but it is unsuitable for immediate post-earthquake traffic flow modelling because drivers lack the information on traffic congestion needed to select appropriate alternative routes or may panic as a result of the chaos following the earthquake (Pel et al., 2011) .

Up to now, no approach has been developed that provides a comprehensive assessment of traffic patterns and congestion formulation in a transportation network immediately following an earthquake. When a damaging earthquake occurs, the road capacities throughout the city are impaired, and traffic patterns change abruptly and continuously. These changes, combined with the traffic information unavailability and driver panic, present a challenge to real-time traffic modelling ("real time" indicates the time-varying characteristics of congestion condition, traffic flow and traffic information after the earthquake). To address this issue, this paper presents an agent-based framework to model the time-varying traffic patterns in an urban region immediately following an earthquake. This model accounts for the abrupt changes in traffic patterns that result from impaired road capacity, changes in drivers' routes, unavailability of traffic information and driver disorientation caused by the stress caused by the sudden occurrence of the earthquake. The approach is illustrated by modeling traffic patterns in Tangshan City, China to study the mechanisms of congestion formation and to evaluate the performance of that transportation network following a severe earthquake. The post-disaster performance metric is the ability of the traffic system, in its damaged condition, to permit injured people to be transported to hospitals within a reasonable period of time (Brodsky, 1990).

## 2 Methodology

Traffic modelling is aimed at estimating OD demands between traffic analysis zones (TAZ), and simulating travel behaviors of vehicles in the transportation network. Current methods usually regard daily traffic as an equilibrium system, in which traffic demands and travel delays can be obtained by traffic simulation. However, the sudden occurrence of an earthquake may change the travel demand, disrupt the equilibrium of the traffic system and lead to significant traffic congestion. In this section, we introduce the user equilibrium (UE) model to generate the pre-earthquake traffic; subsequently, we incorporate agent-based model in dynamic traffic assignment (DTA) to simulate the post-earthquake traffic conditions under different emergency scenarios.

### 2.1 Pre-earthquake traffic calculation with static equilibrium model

Prior to the occurrence of the earthquake, normal traffic conditions are analyzed using the UE model in two steps: traffic generation and traffic assignment.

### 2.1.1 Traffic generation

Traffic generation estimates the number of trips that originate from and arrive in each TAZ. Generation consists of trip production and trip attraction. In traffic generation, a given network first is separated into several TAZs, and then the gravity model (Chatterjee and Venigalla, 2004), which is by far the most widely used technique for trip distribution, is applied to capture the main traffic demand. The gravity model assumes that the number of trips between locations $i$ and $j$ per unit time, $T_{ij}$, is proportional to some power of the population of the origin ($m_i$) and destination ($n_j$) locations, and decays with the distance between them, $r_{ij}$,

$$T_{ij} = \frac{m_i^g n_j^h}{f(r_{ij})} \tag{1}$$

where $g$ and $h$ are adjustable exponents and the distance function $f(r_{ij})$ is chosen to fit the empirical data. Occasionally $T_{ij}$ is interpreted as the percentage of travelers travelling from *i* to *j*, or an effective coupling between the two locations (Krueckeberg and Silves, 1974). This model is widely used both in normal and post-hazard traffic conditions (Moriarty et al., 2007; Chang et al., 2012).

**2.1.2 Traffic assignment**

After the traffic demand is generated by the gravity model, it is loaded into a given transportation network. The travel time, the route choice of each traveler and the traffic flow of each road, is then calculated by the static UE model. The static UE model follows the *Wardrop Principle* (Wardrop 1952), in which no user can switch routes within the network while travelling to enhance his/her travel time. The traffic assignment so defined can be solved as an optimization problem. To facilitate the optimization, the static UE model is defined mathematically by (Beckmann et al., 1956):

$$\min Z(x) = \sum_a \int_0^{x_a} t_a(w) dw \tag{2}$$

with the following constraints:

$$\sum_{k \in \Psi_{rs}} f_k^{rs} = T_{rs}, \forall r \in R, \forall s \in S \tag{3}$$

$$f_k^{rs} \geq 0, \forall k, r, s \tag{4}$$

$$x_a = \sum_r \sum_s \sum_k f_k^{rs} \delta_{a,k}^{rs}, \forall k, r, s \tag{5}$$

in which $x_a$ is the traffic flow volume on link *a*; $t_a$ is the travel time on link *a*, which is an impedance function recommended by the Federal Highway Administration (Branston, 1976), given by Eq (6) below; $f_k^{rs}$ is the traffic flow volume of the $k^{th}$ route between origin *r* and destination *s*; *R* and *S* are the sets of all the origins and destinations in the traffic network, respectively; $\delta_{a,k}^{rs}$ is a Boolean function showing whether the $k^{th}$ route between $r$ and $s$ passes link $a$; $\psi_{rs}$ is a set of all the routes between $r$ and $s$.

The impedance function is defined as (Branston 1976):

$$t_a(x_a) = t_0(1 + \alpha \left[\frac{x_a}{C_0 \cdot N_L}\right]^\beta) \tag{6}$$

where $t_0$ is the free-flow (unimpeded) travel time of link $a$; $C_0$ is the traffic carrying capacity of one lane; $N_L$ is the number of lanes of link a; $\alpha$ and $\beta$ are variable parameters of the impedance function. The total time cost of the $k^{th}$ route between $r$ and $s$, $C_k^{rs}$, can be determined by:

$$C_k^{rs} = \sum_a t_a(x_a)\delta_{a,k}^{rs}, \forall k, r, s \tag{7}$$

The objective function defined by Eq. (2) minimizes the sum of the integrals of the total time cost of all links by a set of link flows subjected to the flow conservation conditions (Eqs. 3 and 5), which requires that all the traffic demands are assigned within the traffic network. Eq (4) implies that the traffic flow on each route is non-negative. Finally, Eq (7) determines the total time cost of each route.

The pre-earthquake traffic modelling with the UE model generates a steady-state traffic flow before the occurrence of earthquake, which serves as the initial condition for the post-earthquake traffic flow simulation. The earthquake is assumed to occur suddenly without warning, and the post-earthquake simulation is described next.

**2.2 Post-earthquake real-time traffic simulation with agent-based model**

Agent-based modelling (ABM) is a powerful technique for the analysis of distributed complex systems (Niazi and Hussain 2011). ABM is suitable for system modelling under three conditions: (1) The problem domain is spatially distributed; (2) the subsystems exist in a dynamic environment; and (3) the subsystems need to interact with each other with more flexibility (Adler and Blue 2002). The dynamic traffic assignment (DTA) is well suited to an agent-based approach because of its spatially distributed nature and its alternating busy-idle operating characteristics (Wang, 2008), and it has been applied to model the behavior of drivers under congested traffic conditions (Hidas, 2002; Silva et al, 2006). Furthermore, an ABM is convenient for handling the complexities of human behavior and decision-making when confronted with a non-familiar situation (Panwai and Dia 2007), e.g., the driver may panic as a result of the chaos following the earthquake and be disoriented, thus it was selected in this paper for the emergency traffic simulation immediately after the earthquake.

This study employs two fundamental agents: link and driver (car). With the pre-earthquake traffic information available from the UE model, the post-earthquake traffic simulation starts with the initialization of the ABM, as described below, and then dynamically models the driver behavior and route choices under emergency condition.

### 2.2.1 Initialization of ABM

With the steady-state traffic information defined from the UE model and obtained through the two-step calculation, including travel time and traffic flow volume of each link, the number of cars on each link, $I_a$, is:

$$I_a = x_a \cdot t_a \tag{8}$$

The probability of an individual car on link *a* having a specific OD, *r-s*, is:

$$\Pi_{a,rs} = \frac{\sum_{k \in \psi_{rs}} f_k^{rs} \delta_{a,k}^{rs}}{x_a} \tag{9}$$

With Eq (9), each car on link *a* is assigned an OD randomly in the modelling. The probability of an individual car with an assigned OD, *r-s*, having a specific route, *rt*, is:

$$\Pi_{a,rt}^{rs} = \frac{f_{rt}^{rs} \delta_{a,rt}^{rs}}{\sum_{k \in \psi_{rs}} f_k^{rs} \delta_{a,k}^{rs}} \tag{10}$$

With Eq (10), each car on link *a* is assigned a route randomly in the modelling. Assuming cars are evenly distributed on the link, the time for the $i^{th}$ car in the queue to arrive at the intersection between links equals:

$$t_{i,a} = t_a(x_a) \cdot \frac{i}{I_a} \tag{11}$$

With the above models in Eqs. (8) to (11), the traffic modelling results with static UE model are converted to the information associated with the agents (every link and every car) in the network. The associated information is listed in Table 1.

Table 1. Information associated with each agent

| Agent type | Information |
|---|---|
| Link | Link ID; Number of cars; IDs of cars on each link; Length; Number of lanes; Traffic capacity ($\alpha$ and $\beta$ in Eq.6) |
| Car (driver) | Car ID; ID of link the cars is on; Origin-Destination; Time to the cross road; Route; Mental state; ID of car ahead |

This information provides the initial conditions for the following agent-based modelling.

### 2.2.2 Capacity modelling of damaged road network

Bridges, buildings and other facilities may incur significant damage during a severe earthquake. The traffic capacity of a road network is impaired mainly due to two reasons: the damage of bridges (river or road crossings) and debris from damaged infrastructure spilling onto the roadway (Stern and Sinuay-Stern, 1989). In this study, buildings and bridges both are considered vulnerable to earthquake. Road damage may also be incurred by the earthquake, its effect is not considered in this study because the traffic capacity is more affected by bridge damage in existing earthquake observations (Chen et al., 2008).

To estimate the impaired bridge traffic capacity following the earthquake, the bridge seismic fragility is considered (Padgett and DesRoches, 2007). In this paper, the bridge fragility is defined as the probability of exceeding a prescribed limit state $LS_i$ conditioned on peak ground acceleration (PGA):

$$P(LS_i|PGA = y) = \Phi(\frac{\ln y - \lambda_i}{\xi_i}) \qquad (12)$$

where $\Phi(\cdot)$ is the cumulative distribution function of the standard normal random variable, and $\lambda_i$ and $\xi_i$ are the parameters (logarithmic mean and logarithmic standard deviation) of the lognormal distribution, respectively, for the $i^{th}$ damage state. According to the seismic capacity survey of existing bridges, the fragility parameters defining the boundaries between four damage states of multi-span simply supported reinforced concrete bridges are described in Table 2 (Wu et al. 2014).

**Table 2 Fragility parameters for multi-span simply supported reinforced concrete bridges (g)**

| Damage state | Slight | | Moderate | | Severe | | Complete | |
|---|---|---|---|---|---|---|---|---|
| Parameters | Median | Dispersion | Median | Dispersion | Median | Dispersion | Median | Dispersion |
| | 0.45 | 0.96 | 0.56 | 0.96 | 0.71 | 0.96 | 0.98 | 0.96 |

We assume, similar to the approach taken by Padgett and DesRoches (2007), that in the state of emergency immediately following the earthquake, drivers stop using the bridge *only* if significant damage is visible, i.e., severe or complete damage has occurred. Accordingly, the traffic capacity is assumed to remain unchanged until severe damage occurs, at which time the bridge traffic capacity decreases to 0.

Earthquake-induced damage to structural systems, non-structural components and building contents may cause debris that impairs roadway traffic capacity (Ellidokuz et al 2005). This debris may block a portion of the roadway and force vehicles to weave into another lane to avoid it. The traffic capacity of such a weaving area, $C_{WA}$, has been studied (Roess et al., 2011):

$$C_{WA} = C_0 \cdot r_s \cdot r_{VR} \tag{13}$$

where $r_s$ is the lane correction coefficient, which is 0.3, 1.8, 2.6, 3.4 and 4.0 respectively for the cases of 1, 2, 3, 4, and 5 lanes, $r_{VR}$ is the volume ratio correction coefficient, which decreases with the percentage of weaving volume in total flow volume and is defined in the literature (Roess et all, 2011), and $C_0$ is defined in Eq. (6). In this study, we model the capacity of debris-impacted road by dividing it into free flow length and weaving length. Suppose among the road length *l*, the total length of weaving area is $l_w$; the traffic capacity of the debris-affected road is

$$C_W = C_0 \cdot \frac{r_s \cdot l_w + N_L \cdot (l - l_w)}{l} \cdot r_{VR} \tag{14}$$

The travel time of the debris-affected roadway is calculated by:

$$t_w = t_0 \cdot \frac{c_0}{c_w} \tag{15}$$

Whether a road is affected by building debris also depends on the distance between then road and building, as well as the height of building. The probability that building debris clutters the roadway is larger for taller buildings close to the road than otherwise. Since the mechanism of earthquake-induced building debris is outside the scope of this study, we consider debris effects only if the roadway is within 10m of a building and if the buildings are more than 25m high and are severely damaged by the earthquake, as described subsequently. The probability of debris obstructing one lane of the road is assumed to increase linearly with the height of building - 0% for height of 25m and 100% for height of over 75m, and the length of debris-affected region of the roadway is assumed equal to the frontage of the building along the road.

**2.2.3 Dynamic modelling of agents**

Past earthquake surveys have shown that once an earthquake occurs, most individuals who are on the road change their original destinations, heading for or returning to home immediately, picking up their spouses or children, etc (Liu et al., 2008; Ahn et al., 2014). Social vulnerability to disasters depends on the nature of the affected urban area, and may have a significant impact on the abrupt change of traffic patterns. Since a detailed study of social vulnerability is beyond the scope of this study, however, we assume in the following illustrations that 30% of drivers maintain their original destinations, 40% return to their homes, 20% drive to school to pick up their children and 10% drive to hospitals to meet their injured relatives.

The possibility that a driver behaves irrationally under the stress of an emergency situation is also considered in the ABM. An individual's response to an earthquake depends on several factors including prior earthquake experience, earthquake-related education, the observed damage conditions of surrounding facilities, and whether it is apparent that people have been severely injured or killed (Nigg, 1984). Past observations of people's behavior during disasters have revealed that some may panic and become disoriented for a short period, and may simply follow others to escape (Helbing et al., 2010). Thus, we assume that a small percentage of drivers will respond in this manner immediately after the earthquake, and simply follow the car ahead. These drivers will regain their composure and choose a new route to their destination within 1 to 20 minutes.

The availability of traffic information is important and is also considered in this study. If real-time traffic information is delivered to drivers by GPS or public media, they will plan their route wisely to avoid roads that are congested or closed. On the other hand, if real-time traffic information is not available, drivers will choose their routes based on experience, which may create further traffic congestion and closures. A rational driver will choose his/her route based on the Fermi process (Wu et al., 2010; Patriksson, 2015; Mahnke et al., 2005) once his/her new destination has been identified. Fermi-process was developed from pairwise comparison processes, and has been widely used in evolutionary game theory for decision making (Szabó and Tőke, 1998). Applied in traffic flow modelling , the Fermi-process holds that once the OD is determined, the route with a shorter time will be selected by the driver with a high probability, through the following function (Ben-Akiva and Bierlaire, 1999):

$$\Pi_{rt,i}^{rs} = \frac{e^{-\theta \cdot t_{rt,i}^{rs}}}{\sum_k e^{-\theta \cdot t_{rt,k}^{rs}}} \tag{16}$$

in which $\Pi_{rt,i}^{rs}$ is the probability that the driver with OD of *r-s* will choose route *i*; $t_{rt,i}^{rs}$ is the total travel time from *r* to *s* by route *i*; and $\theta$ is the diversion intensity parameter, which describes the capacity of drivers to predict the travel time of different routes in the model. If $\theta = \infty$, the driver will select the route with the shortest time with probability 1; if $\theta = 0$, the driver will select the route randomly. The value of $\theta$ is 0.1 in the existing literature on dynamic traffic modelling under normal conditions (Nguyen et al., 1988; Tanaka et al., 1998; Lam et al., 1999). Considering the fact that traffic conditions become more unpredictable after the earthquake, the value of $\theta$ is set to be 0.05 following the earthquake in our analysis. The Fermi process assumption generally causes the shorter roads to be associated with heavier traffic; the few exceptions are because of higher mean speed, less congested roads or other benefits.

In the model, the driver's mental state and the traffic condition are updated every minute; the number of cars on each link, as well as other information associated with each car including its position, the O-D pair and the selected route, are updated every minute. The velocity of each car is determined by the traffic condition to the next intersection; according to Eq.(6), the travel time for that distance can be calculated, and then the velocity is obtained. The velocity is updated every 10 seconds for each car during its moving, to be consistent with the changing of congestion level. If the head-to-head space of cars is smaller than 8.5m on a link (the car length is around 6m for ordinary cars), the link is judged to be in a congested state, which will result in the complete blockage of the upstream intersection, as illustrated in Fig.1.

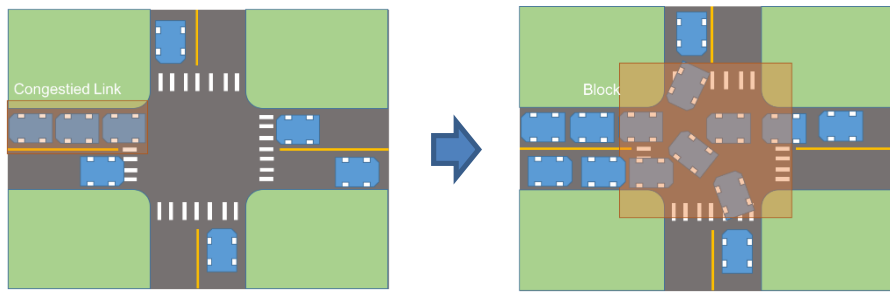

**Fig.1 Blockage of an intersection due to congestion of a link**

Finally, an important feature of this model is that it includes additional traffic flow due to the need to deliver injured people to a hospital or emergency treatment facility. It is assumed 10 additional cars join the transportation network per minute; their origins are randomly distributed in the network and their destinations are the nearby hospitals. When cars reach their destinations, they park there and leave the transportation network. The flowchart in Figure 2 illustrates the entire ABM-based simulation process.

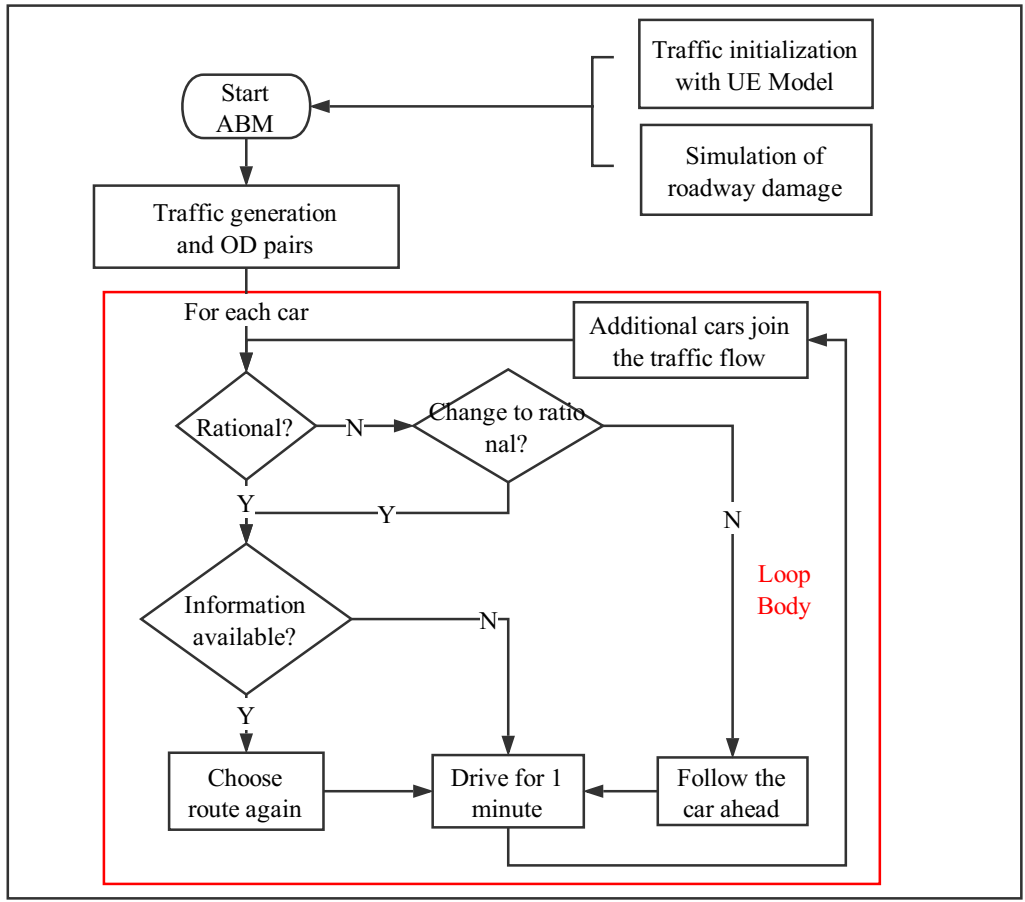

**Fig.2 Flow chart of traffic simulation using ABM**

## 3. Illustration

The post-earthquake traffic flow methodology is illustrated using the transportation network in Tangshan City, Hebei Province in China as a numerical case study. Tangshan City was severely damaged by a moment magnitude $M_w = 7.8$ earthquake occurring in July, 1976, and resulting in a large number of fatalities and great economic losses. The performance of the transportation system immediately after an earthquake is modelled and evaluated in the following.

### 3.1 Tangshan transportation network and earthquake-induced damage

An overview of Tangshan City and the major roads in its surface transportation network are shown in Figure 3a. Tangshan City has an area of around 380 km² and a population of 2,100,000. The transportation network is modeled as consisting of 216 nodes and 750 links. Except the national highway, "G112", whose design speed is 60km/h, the urban streets in the map have design speeds of 25, 30 and 35km/h, respectively, and the parameters in Eq. (6) for these streets are: $\alpha$ = 3.5, $\beta$ = 4.0 for design speeds of 25km/h and 30km/h; $\alpha$ = 4.5, $\beta$ = 4.0 for design speed of 35km/h (CJJ-37). We assume that the bridges are the only roadway transportation infrastructure components that are vulnerable to earthquakes. There are a total of 20 bridges, the positions of which are shown in Figure 3a. The OD travel demands depend on the locations of business zones, residential zones, schools, hospitals and other social support facilities. Figure 3b gives the locations of office buildings, schools and hospitals, as well as the population density of residents in the form of a heat map, from which the OD travel demands are generated for both the pre-earthquake and the post-earthquake traffic modelling. The simulation is performed through MATLAB (2014) programming.

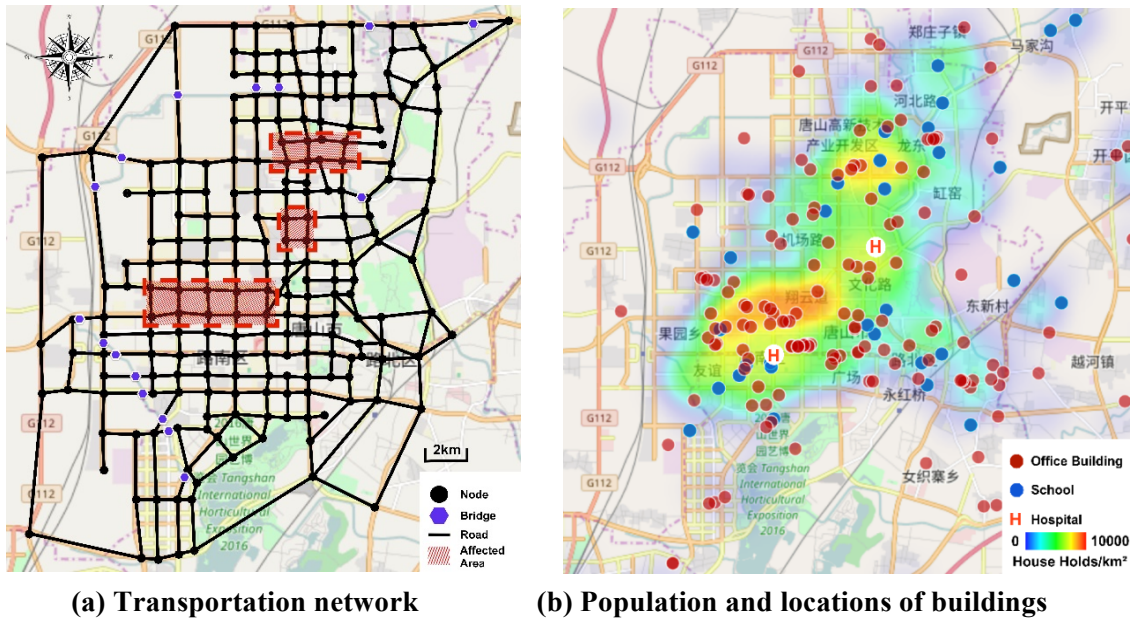

(a) Transportation network     (b) Population and locations of buildings

**Fig.3 Overview of Tangshan City**

For illustration, we assume the occurrence of an earthquake with magnitude of 8 (corresponding to a PGA = 0.25g)[1] (GB50011-2010), comparable to the 1976 earthquake. According to Eq. (12), a bridge has a probability of 15% to be severely damaged. The bridge damage levels would be positively correlated through the use of a common bridge design code in engineering practices, construction conditions, and code enforcement (Vitoontus and Ellingwood, 2013). For simplicity, the correlation coefficient among bridge fragilities is assumed to be 0.7. The relative frequency of the number of bridges severely damaged is obtained by Monte Carlo simulation employing a Gaussian copula (Cario and Nelson, 1997), with a mean value of 3.7. Therefore, 4 bridges are assumed to be severely damaged in a typical scenario of bridge damage in the following simulations.

---

[1] GB5001, the seismic design code for China, stipulates earthquakes for design by magnitude numbers. These magnitude numbers should not be confused with the customary magnitudes, such as moment or body-wave magnitude, in common usage in the seismology community.

A survey of the seismic performance of existing buildings of North China (Zhang et al., 2004) shows that the probability that the building performance level exceeds "severely damaged," when subjected to an earthquake with magnitude of 8, is about 20%; this value was used herein. The correlation coefficient of building damage level due to common design and construction practices is also assumed to be 0.7. Once severe damage occurs to a building, its debris may obstruct one lane of the roadway with a probability depending on the building height (>25m) and its distance from the street (<10m). According to these heights and distances to streets, the possible debris-affected areas are shown in Fig.3a. The results of one simulation of the damaged transportation network are shown in Figure 4; four bridges are severely damaged and lose traffic capacity completely, and 25 streets are affected by debris, as indicated by the weaving length ratios shown in Figure 4. The post-earthquake modelling and evaluation of traffic system are performed on this simulated damaged transportation network, as described in the following section.

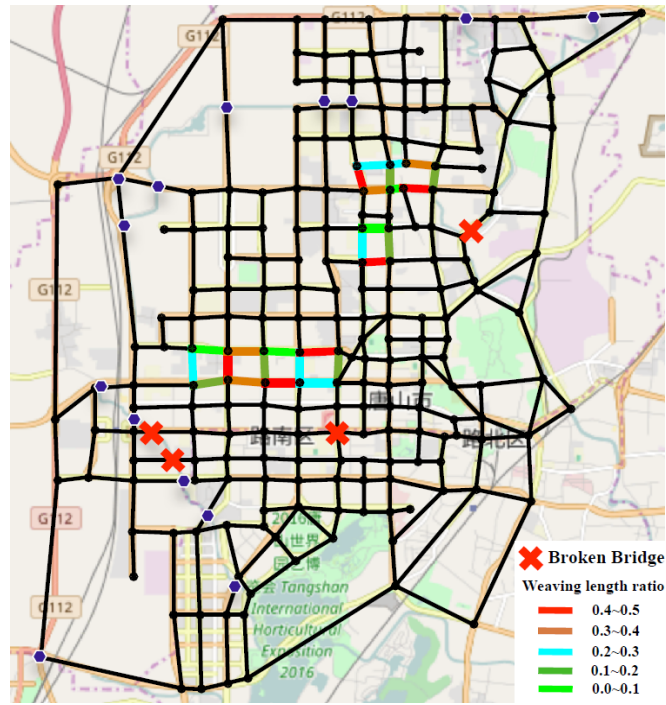

**Fig.4 A scenario of damaged transportation network**

**3.2 Immediate post-earthquake modelling of Tangshan traffic system**

The time of occurrence of the earthquake affects emergency traffic conditions following the earthquake significantly because the traffic patterns and population distribution vary throughout the day. In this example, we assume that the earthquake occurs during the morning rush hour, when approximately 25,000 cars are on the road network. In terms of behavior rationality and information availability, we consider three cases: (1) 30% of drivers panic, and all traffic information is available; (2) 70% of drivers panic, and all traffic information is available; and (3) 70% of drivers panic, and no traffic information is available;

The traffic congestion before the earthquake is obtained by employing the UE model, as illustrated in Figure 5a; the congestion of the road is measured by the travel speed, as identified by the different colors in Figure 5: red (speed is lower than 15km/h, orange (speed is between 15km/h and 25km/h), yellow (speed is between 25km/h and 35km/h) and green (speed is higher than 35km/h). The average travel speed before the earthquake is 24.4km/h, which is consistent with that (24.0km/h) reported in traffic survey (https://bit.ly/2OIgxYa, accessed on 08/03/2018). Figures 5b to 5d illustrate the congestion within the transportation network 10 minutes after the earthquake for the three cases obtained through an ABM-based simulation. A comparison of Figure 5a and 5b reveals how the traffic congestion becomes more significant because the capacity of transportation network is impaired by damages to bridges and the accumulation of building debris on the roadways. The average speed of all cars drops from 24.4km/h to 21.7km/h. A comparison of Figures 5b and 5c shows the effect of drivers' irrational behavior on traffic conditions. As the percentage of drivers that become disoriented increases from 30% to 70%, the congestion level becomes larger, and the average speed drops from 21.7km/h to 20.1km/h. Figure 5d indicates that when traffic information is unavailable, the congestion level increases further, to the point where the average speed is only 16.8km/h. Taken as a whole, Figures 5(a) – 5(d) show that the mental state of drivers and availability of traffic information under emergency conditions both have a significant impact on the congestion within a transportation network above and beyond that caused by structural damage. As a result, drivers may not be able to choose reasonable alternative routes; this inability may aggravate the functionality of the roadway system.

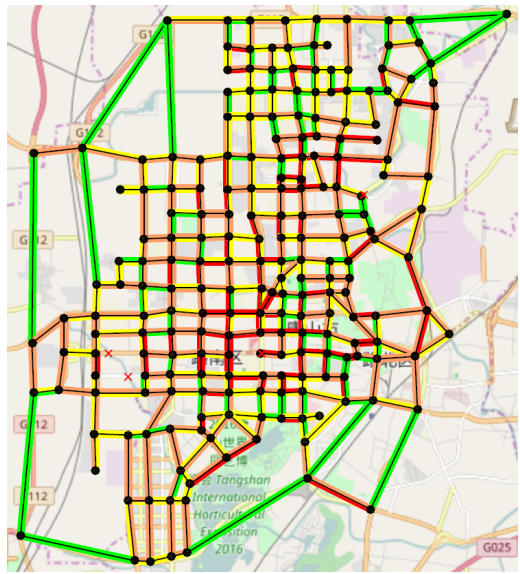
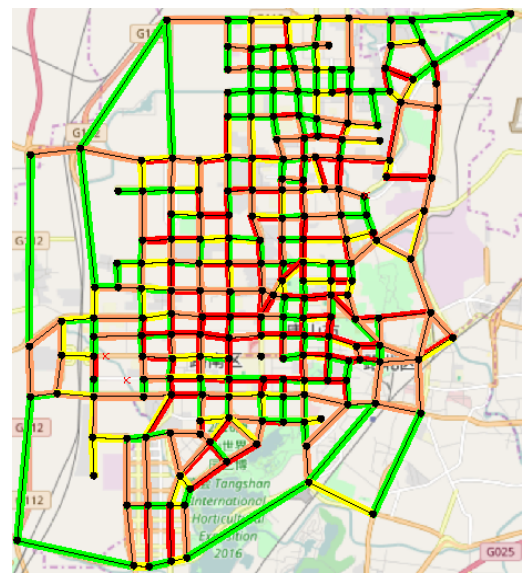

(a) Before the earthquake          (b) 10 min after the earthquake (case 1)

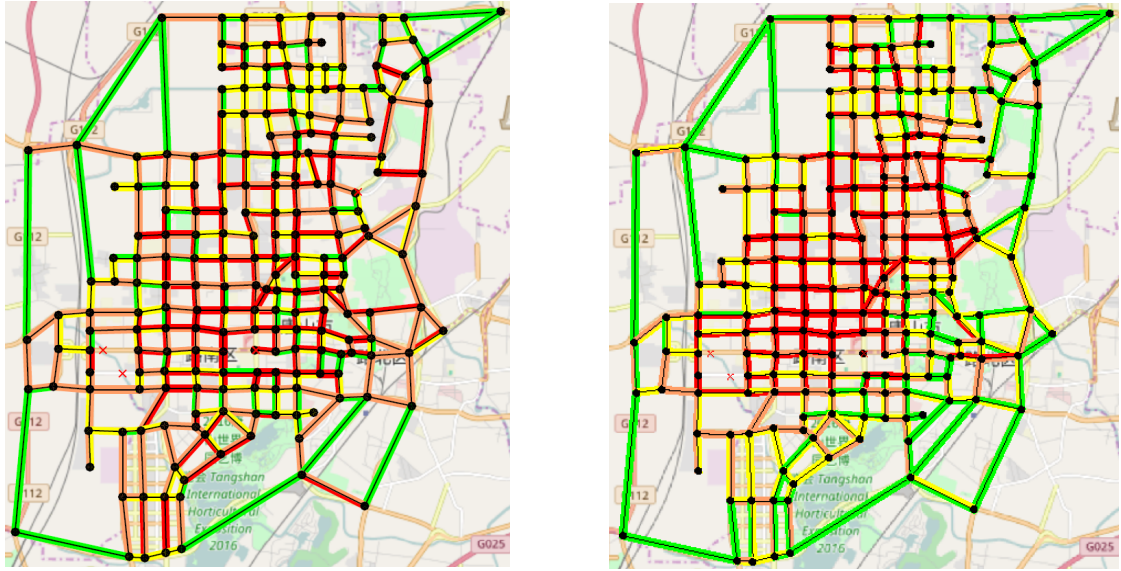

**(c) 10min after the earthquake (case 2)    (d) 10min after the earthquake (case 3)**

**Fig. 5 Comparison of congestion states before and 10 min after the earthquake**

A typical congestion zone is selected for a more detailed examination of the mechanism of congestion aggravation in Figure 5d. This examination is summarized in Figure 6. The space of cars on some links is monitored immediately after the occurring of earthquake. Figure 6 shows that the vehicle spacing on link 1 drops to 2.5 m within 3 minutes of the earthquake due to the influx of cars from link 7, which is blocked due to the earthquake-induced damage to the bridge. The congestion subsequently spreads quickly to links 2 and 3, which are also affected by the influx of cars from link 6, which is completely obstructed, and the spacing of cars on links 2 and 3 drops to 2.5 m following link 1, 4 and 5 minutes after the earthquake, respectively. Links 5 and 4 also reach a congestion state 10 and 13 minutes after the earthquake, respectively, because the intersection at node 2 is blocked. After approximately another 20 minutes (35 minutes after the earthquake), the congestion levels of link 1, followed by links 2, 3, 5 and 4, finally ease.

.

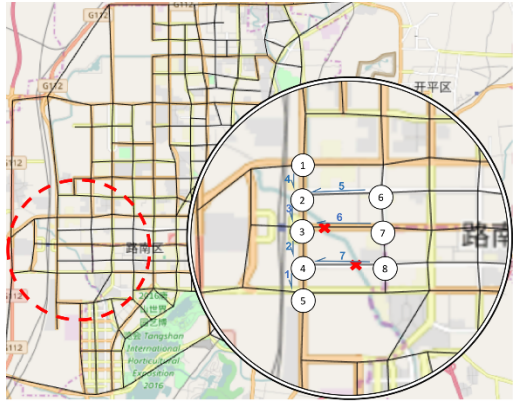 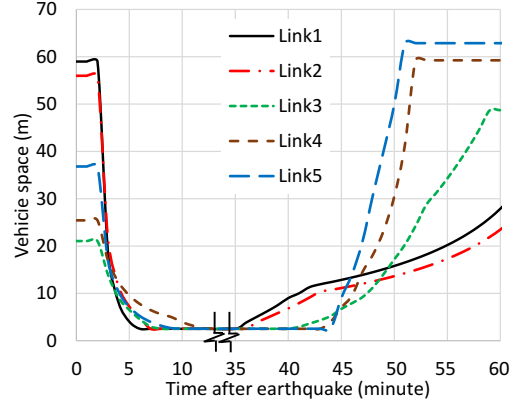

(a) a congestion zone        (b) variation of vehicle space with time

**Fig. 6 Mechanism of congestion aggravation**

## 3.3 Evaluation of the transportation network with regard to emergency medical service

The transportation system must remain operational after the earthquake to facilitate the transportation of injured people to hospitals quickly. Thus, the time required for an injured person to receive emergency medical treatment is a suitable index for evaluating the availability of emergency medical services following the earthquake (Jones and Bentham, 1995; Djalali, et al., 2011). With this in mind, we define the *hospital delivery area*, $A_d$, which is the largest area over which a vehicle can deliver an injured person to the hospital within a specific time, to reflect this capability of the transportation network. The delivery area is time-varying depending on the real-time condition of transportation network. We assume that the delivery area is not dependent on the density of population in the affected area, i.e., that available medical services within the delivery area are sufficient to provided needed services.

To determine $A_d$, the shortest time to reach the hospital from each node is calculated, and the locations on different links that are within a specified time, e.g., 15 min, to the hospital are identified by linear interpolation, as points A, B, C, etc., shown in Fig 7. Hospital delivery area, $A_d$, is defined by connecting these points. For convenience, we now define *hospital delivery radius*, $\rho_d$, by:

$$\rho_d = \sqrt{\frac{A_d}{\pi}} \tag{17}$$

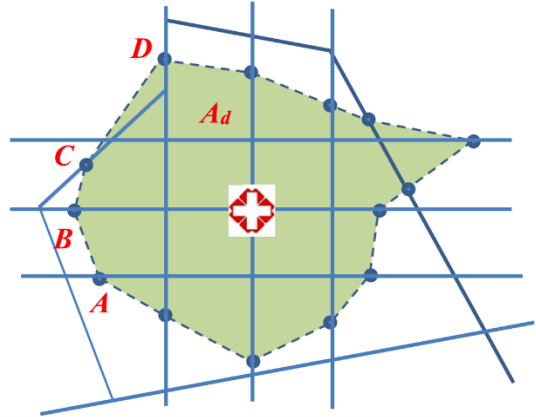

**Fig. 7 Illustrative figure for determining hospital delivering area**

A survey of rescue times in fatal road accidents in the United States in 1990 showed that the rescue time is averages 12 minutes in urban areas and 22 minutes in rural areas (Brodsky, 1990). Seriously injured persons will go into an irremediable state of shock in 15 to 20 minutes. Thus, we define the metrics for emergency medical services by the 15-min delivery areas (or radii) of the two hospitals in Tangshan City during the morning rush hour before the occurrence of earthquake. Figure 8 illustrates this concept; the delivery radii (areas) are 5.73km and 4.12km (104 km$^2$ and 53 km$^2$), respectively.

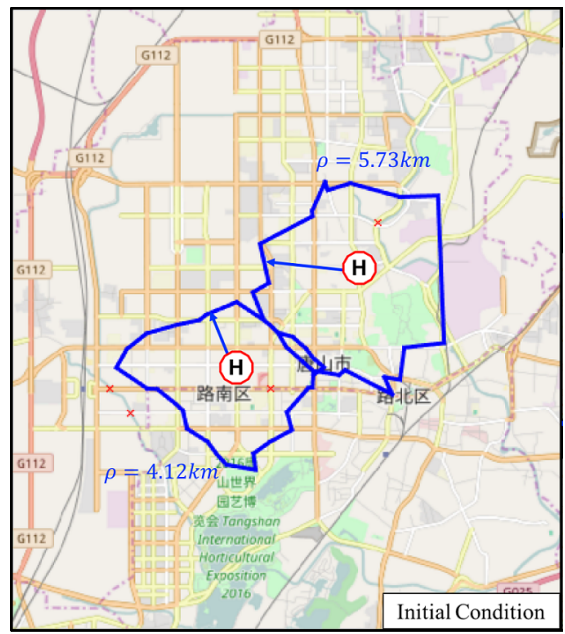

**Fig. 8 The 15-min delivery areas of two hospitals prior to the earthquake**

Following the earthquake, the traffic congestion level becomes more severe immediately, and the 15-min delivery radius suddenly becomes smaller. As the traffic congestion eases, the delivery radius increases gradually. This phenomenon can be seen in Figure 9, which presents the change in 15-min delivery radius with the time after earthquake. Ten minutes after the earthquake, the delivery radius decreases to 4.52km and 2.76km for case 1, 4.22km and 2.69km for case 2, and 3.71km and 2.33km for case 3, showing the effects of physical damage to bridges and buildings, the behavior irrationality of drivers and the unavailability of traffic information. Twenty minutes after the earthquake, all drivers have regained their composure and if traffic information is available (case 1 and case 2), they can choose the most reasonable routes to hospitals and avoid congested roads, and the delivery radii increase significantly. If, on the other hand, traffic information is unavailable, drivers may not be able to avoid congested roads and the delivery radii do not increase as significantly as in cases 1 and 2. Thirty minutes after the earthquake, the congestion level is eased significantly and the delivery radius for cases 1 and 2 is larger than prior to the earthquake; however, the delivery radius for case 3 is still short, implying that the traffic information is very important to ease congestion and the congestion lasts longer if information is not available. It should be noted that this analysis assumes the rational drivers heading to hospitals will change their destinations immediately following earthquake, which ignores the fact that there is usually a time lag between. If the time lag is considered, the congestion level around the hospital will become milder, especially at the early stage, e.g. 10 minutes, after the earthquake.

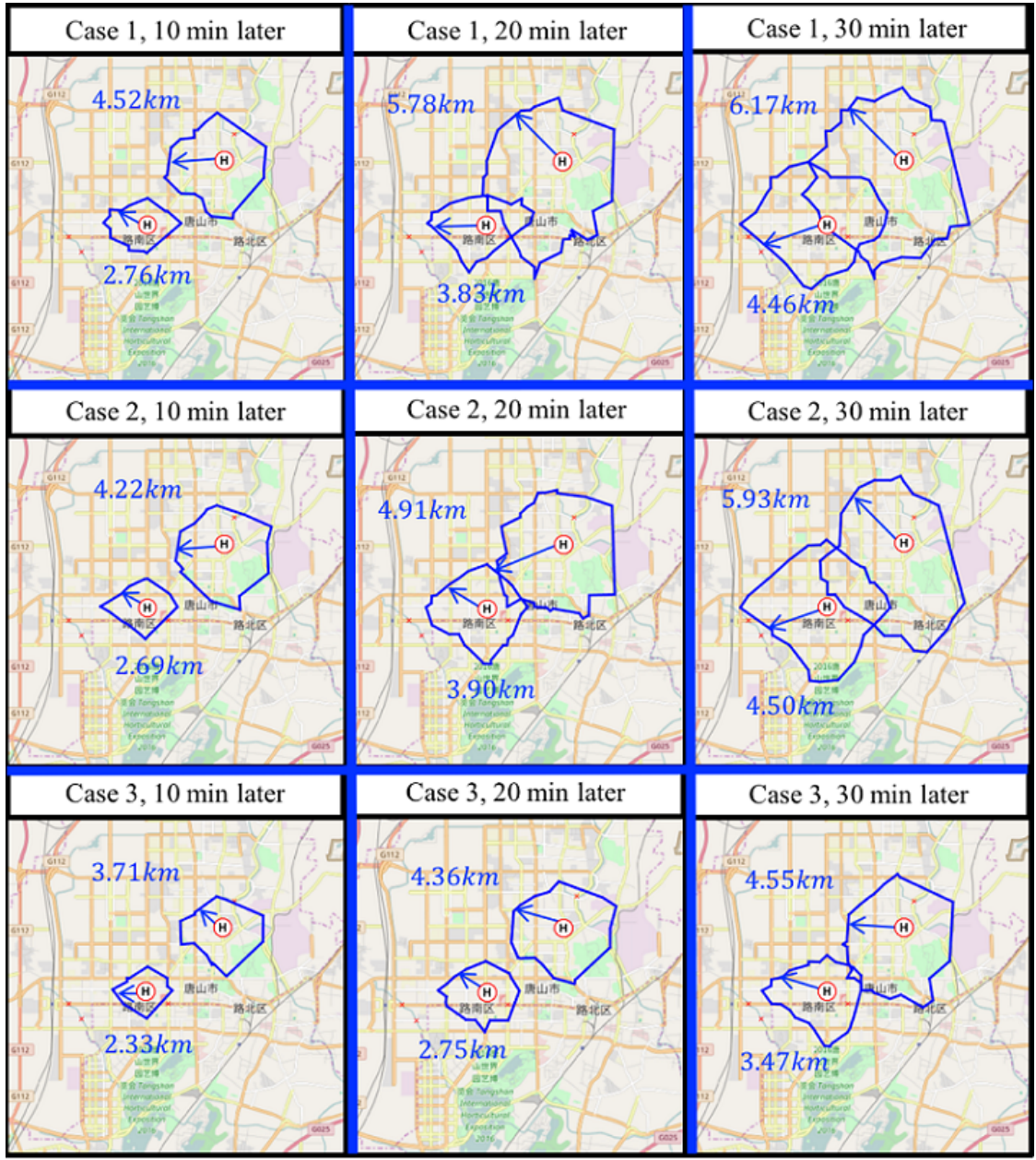

**Fig. 9 Changes in hospital delivery radius with time for the three cases.**

## 4. Conclusions

This paper developed an agent-based methodology to examine the performance of urban transportation networks immediately after a scenario earthquake. The methodology accounts for the unique characteristics of traffic flow under emergency conditions, including the reduction in road capacity due to bridge damage and building debris, the change of driver destination, irrationality of drivers under the stress of emergency conditions, and the unavailability of traffic information. A procedure to simulate the real-time traffic condition was presented, and it was illustrated using the road network in Tangshan City under a hypothetical earthquake during morning rush hour. The performance of the road network was analyzed, and its capability to convey injured people to hospitals was evaluated in terms of the delivery radius of the hospitals.

Our evaluation suggests that (1) the disorientation of drivers and the availability of traffic information have a significant effect on road network congestion a short time after an earthquake occurring during morning rush hour, and the transportation network would be severely impaired by the earthquake; (2) the proposed methodology can capture abnormalities in traffic patterns immediately after the earthquake by accounting for impaired traffic capacity and the effects of emergency responses, and can be used in real-time urban traffic simulation under emergency conditions and to study the mechanism of congestion formulation; and (3) the proposed hospital delivery radius reflects the capability of the transportation network to transfer the injured to hospitals which, combined with the proposed traffic modelling methodology, provides a useful tool for emergency managers to evaluate the performance of transportation systems, and to develop traffic plan for better disaster preparedness and quick response.

A main limitation of this study is that the simulation is time-consuming, and it is difficult to evaluate the post-earthquake performance of transportation system statistically (e.g., the mean and variation of the delivery radius) because a large number of simulations are needed. To overcome this deficiency, an efficient simulation methodology is under development.

**Acknowledgements**

The research described in this paper was supported by The National Key Research and Development Program of China (2016YFC0701404). That support, and additional support from the TNList Chair Professorship, Tsinghua University, and cooperative agreement 70NANB15H044 between the National Institute of Standards and Technology (NIST) and Colorado State University are gratefully acknowledged.

However, the views in this paper represent those of the authors, and do not represent the views of the sponsoring organizations.